\documentclass[aip, amsmath,amssymb, reprint]{revtex4-1}

\usepackage{graphicx}% Include figure files
\usepackage{dcolumn}% Align table columns on decimal point
\usepackage{bm}% bold math
\usepackage[utf8]{inputenc}
\usepackage[T1]{fontenc}
\usepackage{mathptmx}
\usepackage{etoolbox}
\usepackage{color, soul}

\makeatletter
\def\@email#1#2{%
 \endgroup
 \patchcmd{\titleblock@produce}
  {\frontmatter@RRAPformat}
  {\frontmatter@RRAPformat{\produce@RRAP{*#1\href{mailto:#2}{#2}}}\frontmatter@RRAPformat}
  {}{}
}%
\makeatother

\begin{document}

\preprint{AIP/123-QED}

\title[Rapid determination of single substitutional nitrogen N$_s^0$ concentration in diamond from UV-Vis spectroscopy]{Rapid determination of single substitutional nitrogen N$_s^0$ concentration in diamond from UV-Vis spectroscopy}
% Force line breaks with \\
\author{T. Luo}
\author{L. Lindner}%
\affiliation{Fraunhofer Institute for Applied Solid State Physics IAF, 79108 Freiburg, Germany}%

\author{R. Blinder}
\affiliation{Ulm University, D-89081 Ulm, Germany}%

\author{M. Capelli}
\affiliation{School of Science, RMIT University, Melbourne VIC 3001, Australia}%

\author{J. Langer}
\author{V. Cimalla}
\author{F. A. Hahl}
\author{X. Vidal}
\author{J. Jeske}
 \email{jan.jeske@iaf.fraunhofer.de}
\affiliation{Fraunhofer Institute for Applied Solid State Physics IAF, 79108 Freiburg, Germany}%

\date{\today}% It is always \today, today,
\begin{center}
\onecolumngrid
(The following article has been accepted by Applied Physics Letters. After it is published, it will be found at https://doi.org/10.1063/5.0102370)
\end{center}

\begin{abstract}

Single substitutional nitrogen atoms N$_s^0$ are the prerequisite to create nitrogen-vacancy (NV) centers in diamonds.
They serve as the electron donors to create the desired NV$^-$ center, provide charge stability against photo-ionisation, but also are the main source of decoherence.
Therefore, precise and quick determination of N$_s^0$ concentration is a key advantage to a multitude of NV-related research in terms of material improvement as well as applications. 
Here we present a method to determine the N$_s^0$ concentration based on absorption spectroscopy in the UV-Visible range and fitting the 270~nm absorption band.
UV-Visible spectroscopy has experimental simplicity and widespread availability that bear advantages over established methods.
It allows a rapid determination of N$_s^0$ densities, even for large numbers of samples. 
Our method shows further advantages in determining low concentrations as well as the ability to measure locally, which is highly relevant for diamonds with largely varying N$_s^0$ concentrations in a single crystal.
A cross-check with electron paramagnetic resonance (EPR) shows high reliability of our method and yields the absorption cross section of the 270~nm absorption band, $\sigma=1.96\pm0.15$~cm$^{-1}\cdot$ppm$^{-1}$ (in common logarithm) or $\sigma_e=4.51\pm0.35$~cm$^{-1}\cdot$ppm$^{-1}$ (in natural logarithm), which serves as a reference to determine N$_s^0$ concentrations, and makes our method applicable for others without the need for a known N$_s^0$-reference sample and calibration. 
We provide a rapid, practical and replicable pathway that is independent of the machine used and can be widely implemented as a standard characterization method for the determination of N$_s^0$ concentrations.
\end{abstract}

\maketitle

Nitrogen is one of the main impurities in both natural and synthetic diamonds~\cite{ashfold2020nitrogen}.
Related defects, particularly the nitrogen-vacancy (NV) center, attract a broad interest for its special optical and spin properties~\cite{aharonovich2011diamond,doherty2013nitrogen}.
It has been extensively investigated for magnetometry~\cite{degen2008scanning,acosta2009diamonds,rondin2014magnetometry,jeske2016laser,hahl2021magnetic}, bio-sensing~\cite{schirhagl2014nitrogen,wu2016diamond}, nuclear
magnetic resonance (NMR)~\cite{glenn2018high,bucher2019quantum} and scanning probe microscopy~\cite{schell2011scanning,zhou2017scanning}. 
To create NV centers, single substitutional nitrogen atoms, denoted as N$_s^0$, P1 or C-centers, are the prerequisite.
They are doped during the diamond growth or implanted on the diamond surface, followed by irradiation and annealing steps to create vacancies and allow NV centers to form.
They play a role as the typical electron donor to charge the desired NV$^-$ state~\cite{collins2002fermi,fu2010conversion,haque2017overview,hauf2011chemical}, and they are often present 10-100 times more than NV centers even in highly doped material.
They determine the NV charge stability, and optimising the N$_s^0$ to NV$^-$ conversion ratio is crucial for improving the NV center's performance and sensitivity~\cite{luo2022creation}.
Moreover, at parts-per-million (ppm) levels they can act as the main decoherence source of the NV center~\cite{bauch2020decoherence}.
Therefore, knowledge of the N$_s^0$ density gives significant advantages in the choice of material and performance in applications.

To measure the N$_s^0$ density in diamond, the established standard is electron paramagnetic resonance (EPR) spectroscopy~\cite{smith1959electron,van1997dependences,eaton2010quantitative}.
Field modulation is applied to record continuous wave EPR spectra, resulting in a spectrum that approximates the first derivative of the EPR line shape.
Double integration of the acquired spectrum is taken to determine the EPR peak intensity and hence the N$_s^0$ concentration~\cite{tallaire2006characterisation}.
Although this method has been well established, it is often hard to access and labour-intensive, and its high requirement for the surface quality of diamond brings technical difficulties for the measurement.
Moreover, the EPR method measures the spin number in the entire sample volume and averages different areas in the sample.
For diamonds with less homogeneity, e.g. high-pressure, high-temperature (HPHT) diamonds that usually show sectors containing very different N$_s^0$ concentrations in a single crystal~\cite{capelli2019increased}, the spatially resolved determination of the N$_s^0$ density can be of interest rather than the averaged value.
In this regard, optical methods with experimental simplicity and widespread availability have significant advantages for many groups and companies working with NV centers, which makes the Fourier-transform infrared (FTIR) and UV-Visible (UV-Vis) spectroscopy favourable in some cases.  
Depending on the configuration of the spectrometer, both volume and spatially resolved measurements are possible by these optical methods.

The FTIR spectroscopy has been used to determine both N$_s^0$ (at 1130~cm$^{-1}$ and 1344~cm$^{-1}$)~\cite{dobrinets2016hpht} and N$_s^+$ (at 1332~cm$^{-1}$)~\cite{lawson1998existence} centers. 
For this method, a good spectral resolution is required, and different resolutions of the spectrometer have a significant effect on the concentration estimation~\cite{liggins2010identification}. 
Moreover, for diamonds with low nitrogen densities less than a few tenths of ppm, the conventional FTIR spectroscopy often shows insufficient sensitivity in detecting these nitrogen-related centers~\cite{dobrinets2016hpht}.
In comparison, the UV-Vis spectroscopy enables the detection of single nitrogen as low as 0.01~ppm~\cite{de2008determination}, which is much below the detection limit of the conventional FTIR method. 
Furthermore, the UV-Vis spectroscopy with its experimental simplicity can be applied as a rapid method to characterize diamonds, especially for low nitrogen concentrations.

In the UV-Vis spectrum, the absorption band centered at 270~nm has been suggested to determine the N$_s^0$ concentration~\cite{dyer1965optical,chrenko1971dispersed,walker1979optical}.
Extracting the 270~nm band from the spectrum is thus a key step for this method.
%to calculate the N$_s^0$ concentration by UV-Vis spectroscopy.
Different fitting methods have been introduced, but complex spectra and the simplistic nature of the literature methods leads to difficulties to determine the band reliably.
Early on, Sumiya~{\em et al.}~\cite{sumiya1996high} has suggested to subtract the spectrum at 270~nm with the `tail-line'%(Figure~\ref{fig:old}a)
, then calibrate this height with the EPR result. 
The `tail-line' is a straight line fitted with the acquired spectrum at around 600-800~nm. 
This method provides a convenient approach which does not require complicated fitting.
For diamonds without additional spectral components at the `tail' region, it enables a quick and rough estimation of the N$_s^0$ concentration.
When the diamond spectrum shows absorption bands that across the range (for example NV$^-$, NVN$^-$ centers, or the nickel-related broad band centered at 710~nm, etc.), the method will be less accurate and even invalidated.
Especially for chemical vapor deposition (CVD) diamonds, a perfect `tail' without influence by other defects is unusual and difficult to achieve in nitrogen-doped growth.

A more advanced protocol has been introduced by Khan~{\em et al.}~\cite{khan2009charge,khan2013colour}, which avoids being dependent on the `tail' region.
This method relied on more complex fitting components: a `ramp' in the form of $\lambda^{-3}$ well fitting the overall decreasing trend of the absorption spectrum (which can be related to single vacancies~\cite{luo2022creation}, or vacancy clusters~\cite{maki2007effects,jones2009dislocations}); a combination of bands at 360~nm and 520~nm originate from vacancy clusters and NVH$^0$ centers respectively; and a `reference spectrum' including the 270~nm band and its absorption continuum taken from a high-pressure, high-temperature (HPHT) type Ib diamond. The weight of the `reference spectrum' then gives the strength of the 270~nm band and thus the N$_s^0$ concentration. 
This protocol improved the fitting accuracy significantly, it broadened applicable spectrum types as it was independent of absorption bands at 600-800~nm.
Nevertheless, it is still limited by the HPHT reference spectrum that requires a detectable and clear 270~nm band.
Besides, HPHT spectra have their own `ramp' component and potentially other spectral features, using these as reference will not only isolate the 270nm peak but also fit the other components of the reference to the acquired spectrum.
This creates a dependency on the utilised HPHT reference spectrum.
Considering the fundamental difference in material properties between different synthesized diamonds, to avoid using a specific type of reference spectrum that contains the band of interest (270~nm) helps to reduce the fitting uncertainties and further improve the accuracy.

In this work, we present a fitting protocol to estimate N$_s^0$ concentrations reliably and precisely from the 270~nm band in the UV-Vis absorption spectrum. 
This protocol can be more generally adapted for complex spectra, especially for CVD diamonds.
It also avoids the requirement for a reference spectrum with the band of interest and increases the fitting robustness. 
We furthermore calibrate this method with EPR measurements for a series of CVD diamonds with varing N$_s^0$ concentrations.
From the calibration we find the linearity of the two methods and precision of our measurement.
We deduce the absorption cross-section of the 270~nm band for N$_s^0$, enabling the estimation of its concentrations without the requirement of a further reference sample with a known concentration.

We grew six (100) oriented CVD diamonds with varying nitrogen-doping levels for the method calibration.
%The samples were grown in-house with varying nitrogen doping levels.
We pre-characterized their N$_s^0$ concentrations with an EPR spectrometer (Bruker ELEXSYS E580) at room temperature.
The spectrometer was fitted with a Bruker super-high-Q probehead (ER4122 SHQE), and the microwave frequency was set to 9.84~GHz. 
N$_s^0$ concentrations were determined using the built-in spin-counting feature, from the acquisition software (xEPR).
This measurement for N$_s^0$ concentration carries an error of $\sim$6$\%$ (including the statistic error, the fitting error, and the accuracy of the EPR spectrometer). 
For details of the N$_s^0$ concentration see Table~\ref{table:sample}.

\begin{figure}[htpb]
    \centering
    \includegraphics[width=0.48\textwidth]{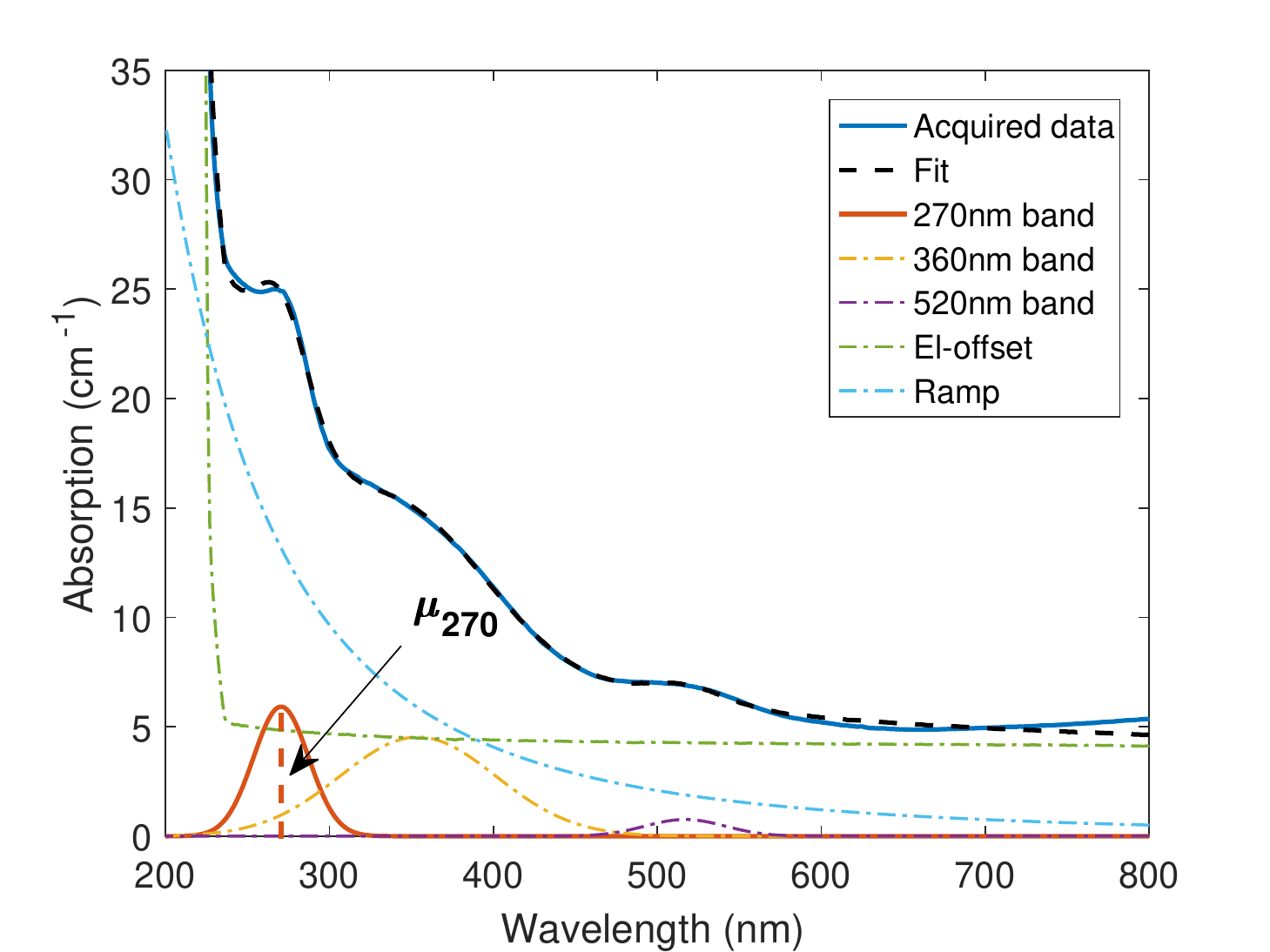}
    \caption{\label{fig:Sp_separation}Fitting for the UV-Vis absorption spectrum including three Gaussian bands at 270~nm, 360~nm and 520nm, a 'ramp' function and an electronic-grade diamond spectrum as the offset. The fitting determines weighting factors to the components. The 270~nm band (red solid curve) can be taken as a direct indication of the concentration of N$_s^0$. }
\end{figure}

To obtain the UV-Vis absorption spectra for the diamond, we measured the diamond transmission $T$ in the range of 200-800~nm using an UV-Vis spectrometer (PerkinElmer Lambda 950) at room temperature, then we deduced the absorption coefficient $A$ following the Lambert-Beer law: $A=$~$-\log_{10}(T)/d$, where $d$ is the sample thickness.
Figure~\ref{fig:Sp_separation} shows a typical UV-Vis absorption spectrum of CVD diamonds, containing three absorption bands centered at respectively 270~nm, 360~nm and 520~nm. We fitted the spectrum with five components (Figure~\ref{fig:Sp_separation}): 

- Three Gaussian functions for the three typical bands in CVD diamonds $g_j(\lambda)=a_j\cdot\exp{(-(\lambda-b_j)^2/(2c_j^2))}$, where $a_j$, $b_j$ and $c_j$ are the fitting parameters for each band $j$=270, 360, 520~(nm). Here 270~nm is the band of interest to extract the N$_s^0$ concentration, 360~nm corresponds to vacancy clusters and 520 nm to NVH$^0$ centers~\cite{khan2013colour}.

- A `ramp' function $r(\lambda)=R\cdot\lambda^{-3}$ (same as in \cite{khan2009charge}), where the factor $R$ is the fitting parameter for this function.

- A spectrum `El-offset', $e(\lambda)$, which is given by an electronic grade diamond (theoretically a `pure' diamond without defect bands in the UV-Vis range). 
This reference spectrum is applied as an offset and baseline to the acquired spectrum, and it is also fitted with a coefficient $d$.

We fitted a sum of the five components to the original spectrum by a non-linear least squares fit: 

$Fit = g_{270}(\lambda)+g_{360}(\lambda)+g_{520}(\lambda)+r(\lambda)+d\cdot e(\lambda)$

Different from previous works, here the key component, the 270~nm feature, is extracted as a Gaussian band. 
For this band (i.e. $g_{270}(\lambda)$ function), a lower- and upper-boundary of the fitting parameter have been introduced for the band-center $b_{270}$ (268-272~nm) and the Gaussian RMS width $c_{270}$ (13-27~nm), in order guide the algorithm to optimize the parameters in a small range. 
The boundaries were selected based on free fitting results of a large amount of samples, they also help to examine whether the fitting outcome exhibits reasonable values: if a final parameter is one of the boundary values, this indicates a unreliable fitting performance, i.e. the boundary has `forced' the algorithm to stop while no optimization has been found in the expected range. 
On the contrary, if the final parameter exhibits a free value in the given range, the fitting was optimized correctly. 
For all our fits this condition was met.

With extracting the 270nm band, we then compare its height, $\mu_{270}$ (in cm$^{-1}$), with the EPR measurements to investigate if the fitting result scales linearly to the EPR and thus can be validated.
In this method, no reference sample with an ideal 270~nm band is required, the only reference spectrum in this method is the `El-offset' component measured from the electronic grade diamond, which is an undoped diamond and thus avoids the variation which occurs in doped reference spectra.
One can easily obtain this spectrum repeatably from any `pure' diamond, or simply use a literature spectrum as the offset for well-calibrated spectrometers.~\footnote{We can provide the spectrum data for the `El-offset' and other technical assistance upon request.}
This reduces the difficulty for implementing the fitting, and greatly avoids introducing undesired spectral components that vary from diamond to diamond.
The `El-offset' keeps the diamond intrinsic spectral feature (with an absorption edge at around 225-235~nm, then being `flat' in the visible range up to 800~nm), instead of using a straight line as the offset.
Importantly, considering that the 270~nm band is located very close to the absorption edge ($\sim$225-235~nm), an additional parameter describing the sharp drop in this regime is necessary to supplement the `ramp' function.
In this sense, introducing the `El-offset' can considerably improve the fitting performance.

For a spectrum with a weak or undetectable 520~nm band, i.e. for samples without NVH$^0$ centers potentially, the fitting method can be also adapted to four components. 
This can help to improve the fitting accuracy for some samples, as the fitting parameters for $g_{520}(\lambda)$ should be nearly zero in this sense, subtracting this band in the fitting function reduces unnecessary fitting parameters.

\begin{figure*}[!hbt]
    \centering
    \includegraphics[width=\textwidth]{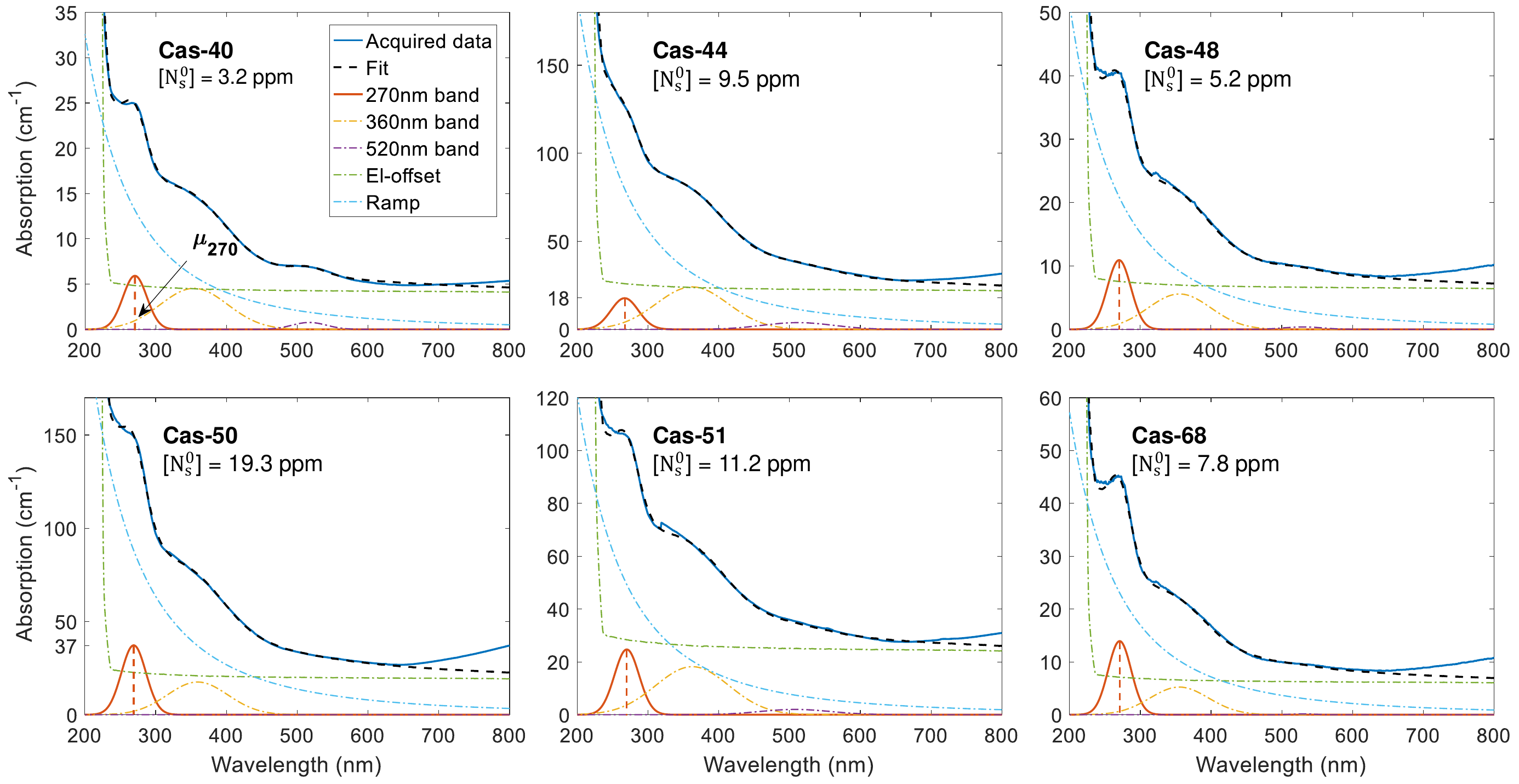}
    % \put(-480,230){\large (a)}
    % \put(-190,230){\large (b)}
    \caption{\label{fig:Sp_all}Fitting result for all six samples. Details of the N$_s^0$ concentration and the peak height of 270~nm see Table~\ref{table:sample}.}
\end{figure*}

Figure~\ref{fig:Sp_all} shows the fitting result for the six samples, they all show a good match between the fitting result and original spectrum.
Mismatches appear at high wavelength ($>$650~nm), which is due to higher spectral features that are not included in the fitting.
A possible candidate for these CVD diamonds can be H2 centers (NVN$^-$) with a zero-phonon line at 986~nm and a broad phonon side band centered at around 800~nm.
For samples with strong features in this higher-wavelength regime (i.e. Cas-48, Cas-50 and Cas-68), a cut-off at 650~nm for the acquired data (i.e. fit for 200-650~nm) can improve the fitting performance.
The good fitting result at lower wavelength shows that our method is independent of higher spectral features.
The spectral variation around 650-800~nm illustrates the problem with defining a `tail' as a baseline reference for the 270~nm peak~\cite{sumiya1996high}.

\begin{figure}[!hbt]
    \centering
    \includegraphics[width=0.48\textwidth]{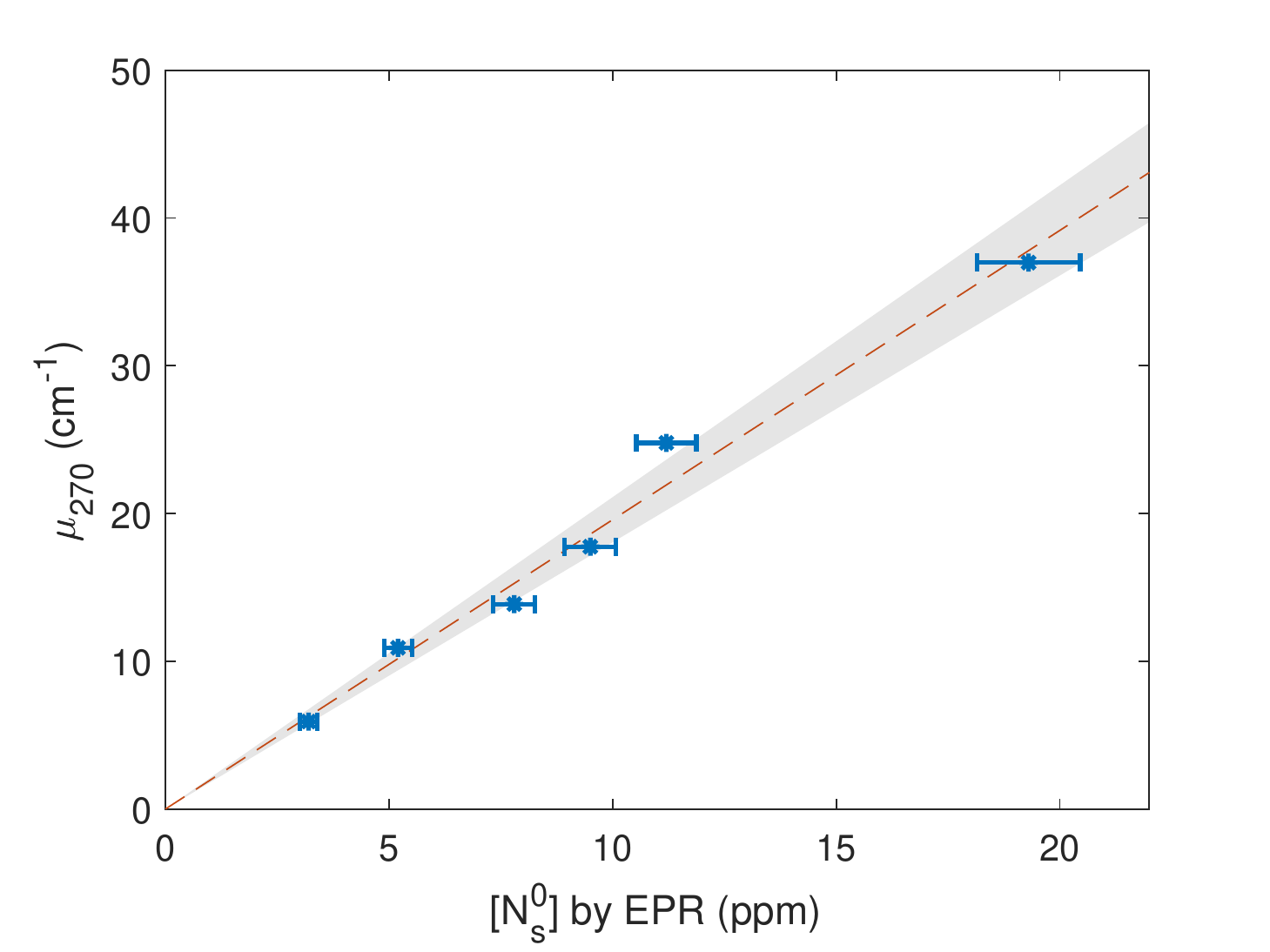}
    \caption{\label{fig:UVVis_EPR}The UV-Vis result by our fitting method is well aligned to the EPR result with a linear correlation. The linear fit shows a slope of 1.96$\pm$0.15~cm$^{-1}\cdot$ppm$^{-1}$ for $\mu_{270}$ given by decadic absorption coefficients. The error of the slope is given by 95\% confidence interval of the fitting parameter (gray area in the plot).}
\end{figure}

As the next step, we compare the height of the extracted 270~nm Gaussian peak, $\mu_{270}$, to the EPR result (Fig.~\ref{fig:UVVis_EPR}).
%They are well aligned to each other, showing a linear correlation.
The good agreement of the two methods shown by the close-to-linear arrangement of the data points, proves the reliability of our fitting method.
The 270nm band has been assigned only to N$_s^0$ and has been used in several papers to determine only the N$_s^0$ concentration~\cite{sumiya1996high,khan2009charge,khan2013colour,edmonds2021characterisation}.
We note however that an earlier research has suggested that the 270~nm band can be influenced by both N$_s^0$ and N$_s^+$, especially for diamonds with low nitrogen concentrations~\cite{jones2009acceptor}, which has never been fully ruled out. 
Although our results can also not fully eliminate the possibility of a contribution of N$_s^+$, the good linear fit of the EPR and UV-Vis results indicates that N$_s^+$ has minor influence if any.
The discrepancies between the two methods in some previous works might have arisen from an inadequate fitting method for the spectrum.
For more details of the band height and the EPR result see Table~\ref{table:sample}. 

From the slope in Figure~\ref{fig:UVVis_EPR}, one can deduce the absorption cross-section $\sigma$ of N$_s^0$ at 270~nm , with the relation: 
\begin{equation}
\label{eq:Ns0}
%   [N^0_s] = \sigma\cdot\mu_{270}
   \mu_{270} = \sigma\cdot[N^0_s]
\end{equation}
where $[N^0_s]$ is the N$_s^0$ concentration in ppm, $\mu_{270}$ is the absorption coefficient of the extracted 270~nm band height (as described above). 
The theoretical derivation of the absorption cross-section is discussed in supplementary material. For common logarithm (i.e. decadic absorption coefficients), the absorption cross-section $\sigma=$~$1.96$$\pm0.15$~cm$^{-1}\cdot$ppm$^{-1}$, the error is given by 95\% confidence interval of the fitting parameter for the linear fit to the two methods.
$\sigma$ can be also given in cm$^2$, i.e. $\sigma=$~$(1.11\pm$$0.09)\times10^{-17}$~cm$^2$, which is converted by multiplying a factor 10$^6m_C/\rho_{dia}$, where $m_C=1.99\times$10$^{-23}$~g is the atomic mass for $^{12}$C and $\rho_{dia}=3.51$~g/cm$^3$ is the diamond density.

For the absorption coefficient calculated by natural logarithm, an absorption cross-section $\sigma_e=$~$4.51\pm$$0.35$~cm$^{-1}\cdot$ppm$^{-1}$ should be applied instead.
In previous research, we have found a single reference of the absorption cross-section at 270~nm~\cite{dobrinets2016hpht}, which has been stated as $1/\sigma=0.6\pm0.1$~ppm/cm$^{-1}$ (i.e. $\sigma=1.67\pm0.28$~cm$^{-1}\cdot$ppm$^{-1}$).
However, the method of its determination, specifically how the spectrum has been separated was not stated.
We thus assume our measurement is more precise, the values of these two completely independent measurements are remarkably close.

From our protocol, one can estimate $[N^0_s]$ directly from their UV-Vis spectra, without calibrating by other methods.
This can be achieved for any samples with both sides polished in the following way: Firstly measure the UV-Vis transmission in percentage then convert it into the absorption spectrum in cm$^{-1}$; secondly separate the spectrum to extract the 270~nm band and obtain $\mu_{270}$ according to the method described above; finally calculate the N$_s^0$ concentration using Equation~\eqref{eq:Ns0} with the absorption cross-section $\sigma$ or $\sigma_e$ (depending on the logarithm type when deducing the absorption from transmission).
The method is setup-independent and applicable for UV-Vis spectra taken by any machine, whether by volume measurements or spatially resolved measurements.
No further EPR/FTIR measurements are required, as Equation~\eqref{eq:Ns0} gives the absolute value of the N$_s^0$ concentration and our value for sigma can be used.
% The method can potentially determine low N$_s^0$ concentrations down to 0.01~ppm for a typical 300~$\mu$m diamond plate, details are discussed in the supplementary material.}

\begin{table}[!hbt]
\caption{\label{table:sample}Details of the samples and results. The N$_s^0$ concentration [N$_s^0$] was measured by EPR, which brings an error of around $\pm$6~\%. The fitting for $\mu_{270}$ has an error of $\lesssim$1~\%.}
\begin{ruledtabular}
\begin{tabular}{ccc}
Sample & [N$_s^0$] (ppm) & $\mu_{270}$ (cm$^{-1}$) \\
\hline
Cas-40 & 3.2 & 5.9\\
Cas-44 & 9.5 & 17.7\\ 
Cas-48 & 5.2 & 10.9\\ 
Cas-50 & 19.3 & 37.2\\ 
Cas-51 & 11.2 & 24.7\\ 
Cas-68 & 7.8 & 13.9\\ 
\end{tabular}
\end{ruledtabular}
\end{table}

%\section{Conclusion}
We developed a fitting method for determining N$_s^0$ concentration via UV-Vis absorption spectrum, which is more widely accessible and easily implementable than EPR measurements.
We showed that the fitting of the identified bands is a reliable way to extract the 270~nm band from the background, as seen by the good match between measurement and fit, and confirmed by the linear relationship between EPR result and our method.
The good agreement with EPR furthermore confirms the assumption that the 270~nm band indeed is mainly caused by N$_s^0$.
Our fitting method performs well for diamond spectra without complex components overlapping with the the fitting components. In other words, it can be widely applied to different diamond types apart from as-grown CVD diamonds.

Furthermore, we deduced the absorption cross-section $\sigma=$~$1.96\pm$$0.15$~cm$^{-1}\cdot$ppm$^{-1}$ (for common logarithm) and $\sigma_e=$~$4.51\pm$$0.35$~cm$^{-1}\cdot$ppm$^{-1}$ (for natural logarithm), which can serve to rapidly determine N$_s^0$ densities from UV-Vis measurements without the need to calibrate a setup via EPR.
This also enables the determination of N$_s^0$ concentrations in a lower end that is hardly detectable by EPR or FTIR methods.
The detectable range of N$_s^0$ concentration is limited by the sensitivity of the UV-Vis spectrometer. 
Based on our spectrometer, we estimated a detectable range for typical 300~$\mu$m diamond plates from 0.01~ppm to 30-50~ppm (details are discussed in supplementary material). 
For thinner/thicker samples, the determination of higher/lower N$_s^0$ concentrations are also possible. 
The range can be expanded by orders of magnitude with more sensitive spectrometers.
The fitting protocol combined with the calibrated absorption cross-section provide a rapid, easy and replicable pathway for the standardized determination of N$_s^0$ concentrations for future research.

\section*{Supplementary Material}
See supplementary material for the discussion of the detectable range of N$_s^0$ concentration and the theoretical derivation of the absorption cross-section.

We thank Brant Gibson, Andrew Greentree, Peter Knittel, Christian Giese, Christoph Schreyvogel and Oliver Ambacher for valuable discussions.
We thank Fedor Jelezko for valuable discussions and supports of the EPR theory.
We thank Michael Ardner, Christine Lell and Michaela Fritz for preparing diamond plates, Dorothee Luick for the technical support of UV-Vis measurements.
T.L and J.J acknowledge the funding by the German federal ministry for education and research Bundesministerium für Bildung und Forschung (BMBF) under Grant No. 13XP5063.
M.C. acknowledges funding from the Asian Office of Aerospace Research and Development (AOARD, funding FA2386-18-1-4056).

The data that support the findings of this study are available from the corresponding author upon reasonable request.

\nocite{*}
\bibliography{aipsamp}% Produces the bibliography via BibTeX.

\end{document}